# Invariant Manifolds and Collective Motion in Many-Body Systems

T. Papenbrock*,† and T. H. Seligman*,**

*Centro Internacional de Ciencias, Cuernavaca, Mexico
†Physics Division, Oak Ridge National Laboratory, Oak Ridge, TN 37831, USA
**Centro de Ciencias Físicas, University of México (UNAM), Cuernavaca, Mexico

**Abstract.** Collective modes of interacting many-body systems can be related to motion on classically invariant manifolds. We introduce suitable coordinate systems. These coordinates are Cartesian in position and momentum space. They are collective since several components vanish for motion on the invariant manifold. We make a connection to Zickendraht's collective coordinates and also obtain shear modes. The importance of collective configurations depends on the stability of the manifold. We present an example of quantum collective motion on the manifold.

## INTRODUCTION

Interacting many-body systems such as atomic nuclei display regular and collective motion as well as complex and chaotic behavior. The interplay of chaotic motion and collective behavior is of particular interest and has been studied for many years. Several authors have investigated models which exhibit chaotic single-particle dynamics in slowly oscillating mean-field potentials [1, 2, 3, 4, 5, 6, 7]. While these models may yield damping and equilibration of the collective mode, they neglect the residual two-body interaction. For the special case of attractive, billiard-like two-body interactions the many-body aspects of collective motion can be studied within the framework of classical dynamics [8].

Here we present an alternative and more general approach to the problem. It is based on the observation that any rotationally invariant system of identical interacting particles possesses low dimensional invariant manifolds in phase space [9]. On such manifolds, the classical motion displays largely collective behavior and decouples from more complex single-particle behavior. The importance of a given invariant manifold depends crucially on its stability properties. If the manifold under consideration is sufficiently stable in transverse directions, the quantum system may exhibit wave function scarring [10, 11, 12] or display a strong revival for wave-packets localized to the vicinity of the manifold [13]. These findings may be directly associated with the slow decay of collective motion despite of the coupling between collective and single-particle motion.

Suitably adapted coordinates for motion on the invariant manifold separate single-particle motion from the collective motion on the manifold [14]. For some types of collective motion, the adapted coordinates can be related to collective coordinates introduced by Zickendraht [16] about thirty years ago. However, the natural coordinates for invariant manifolds are capable of more complicated collective motion such as shearing modes [15].

This contribution is divided as follows. First, we define invariant manifolds in interacting many-body systems and introduce suitable coordinates. Second, we make a connection with the Zickendraht coordinates. Third, we present an example and show that quantum collective motion decays slowly close to invariant manifolds which are not too unstable. Finally we give a summary.

## INVARIANT MANIFOLDS AND ADAPTED COORDINATES

Consider rotationally invariant systems of $N$ identical particles in $d$ spatial dimensions ($d = 2$ or $d = 3$). The Hamiltonian is invariant under both, the action of the rotation group O($d$) and the group of permutations S$_N$. One may now take a finite subgroup $\mathcal{G} \subset$ O($d$) with elements $g$ and properly chosen permutations $P(g)$ such that

$$gP^{-1}(g)(\vec{p},\vec{q}) = (\vec{p},\vec{q}), \quad \forall g \in \mathcal{G} \qquad \vec{p} \equiv (p_1,\ldots,p_{Nd}), \vec{q} \equiv (q_1,\ldots,q_{Nd}) \quad (1)$$

for points $(\vec{p},\vec{q})$ on some invariant submanifold in phase space. On such a manifold, the action of the rotations $g \subset \mathcal{G}$ can be canceled by permutations. These permutations clearly form a subgroup isomorphic to $\mathcal{G}$.

Fig. 1 shows a configuration of four particles in two spatial dimensions that corresponds to a point on an invariant manifold. The operations of elements from the discrete symmetry group $\mathcal{G} = C_{2v}$ can be undone by suitable permutations. This leads to a collective motion with two degrees of freedom which will be shown to be vibrations.

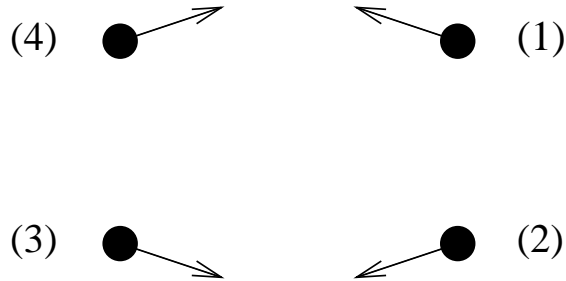

**FIGURE 1.** Collective configuration on invariant manifold. Positions are indicated by filled circles and momenta by arrows.

Fig. 2 shows two spatial configurations of eight (2a) and six (2b) particles, respectively which display a $D_{4h}$ symmetry. If initial momenta display the same symmetry the motion on the invariant manifold will have two degrees of freedom. For eight particles the radii of the two circles will oscillate synchronously, and the two circles will vibrate against each other. For the six particles we will have a vibration of the radius of the circle and of the two particles along the vertical axis. We may choose initial momenta to reduce the symmetry group to $C_{4h}$ which will allow rotations around the vertical axis and thus add an additional degree of freedom. For eight particles we could alternatively choose initial conditions that are limited to a $D_4$ symmetry. Besides the vibrations discussed above this allows for a shearing motion of the two circles thus yielding again three degrees of freedom. Further reductions of symmetry will yield different invariant manifolds with varying degrees of freedom. We will see this exemplified by explicit construction of coordinates.

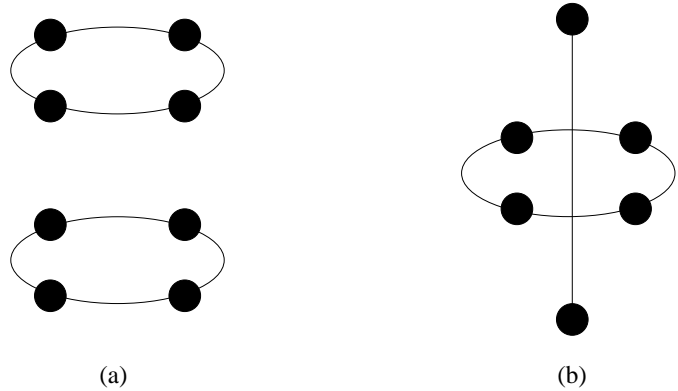

**FIGURE 2.** Configurations of eight (a) or six (b) particles in three dimensions that correspond to invariant manifolds. Positions are indicated by filled circles.

We may use the definition (1) directly for the construction of coordinate systems where non collective coordinates vanish for motion on the invariant manifold. For this purpose we consider the many-body system in Cartesian coordinates in momentum and position space. In what follows we will consider orthogonal transformations in configuration space only; momenta will be subject to the same transformation.

In a Cartesian coordinate system each element $g \in \mathcal{G}$ and each permutation $P(g)$ can be represented by an orthogonal matrix $\mathbf{M}_g$ and $\mathbf{P}_g$ of dimension $Nd$. It is clear that the products $\mathbf{M}_g \mathbf{P}_g^T$ form a matrix group $\mathcal{H}$ that acts onto position and momentum space, respectively. The construction of the coordinate system is now straightforward. Every vector $\vec{p}$ and $\vec{q}$ may be expanded in basis vectors of the irreducible representations (IRs) of $\mathcal{H}$ by means of projectors [17].

$$\Pi_\nu = \sum_{g \in \mathcal{G}} \chi_g^{(\nu)} \mathbf{M}_g \mathbf{P}_g^T. \qquad (2)$$

Here $\chi_g^{(\nu)}$ denotes the character of $g$ in the $\nu$'th IR. The projection onto the identical IR defines the invariant manifold. Note that the identical representation is one-dimensional while the invariant manifolds of interest typically have higher dimensionality. We can find independent vectors on the manifold by projecting from different vectors, but in practice the construction of the independent vectors seems to be unproblematic as we shall see in the example.

## ZICKENDRAHT'S COORDINATES AND INVARIANT MANIFOLDS

About thirty years ago Zickendraht [16] introduced a set of collective coordinates to describe nuclear vibrations and rotations, as well as their coupling with single particle motion. We shall compare these coordinates correspond to the ones we introduced in the previous sections. On one hand this will allow to identify certain vibrational modes of a many-body system with invariant manifolds. On the other hand we shall see that our procedure proposes collective movements that are not of the type described easily in Zickendraht's coordinates.

Following Zickendraht [16] we write the coordinates $\vec{r}_i$ of the $i^{th}$ particle in the center of mass system as

$$\vec{r}_i = s_{i1}\vec{y}_1 + s_{i2}\vec{y}_2 + s_{i3}\vec{y}_3, \quad i = 1, \ldots, N \quad (3)$$

where the $\vec{y}_i$ span the inertia ellipsoid and $s_{ik}$ are called single-particle coordinates. The newly introduced coordinates $\vec{y}_i$ and $s_{ij}$ are not independent. The constraints are $\vec{y}_i \cdot \vec{y}_j = y_i y_j \delta_{ij}$, $i,j = 1,2,3$; $\sum_{i=0}^{N} s_{ij} = 0$, $j = 1,2,3$, and $\sum_{i=0}^{N} s_{ij} s_{ik} = \delta_{jk}$, $j,k = 1,2,3$. The first six equations ensure the orthogonality and normalization of the principal axis of the inertia ellipsoid whereas the next three equations fix the origin at the center of mass system. The last six equations are orthogonality relations of the single-particle coordinates. In the center of mass system, one may therefore characterize the $N$-body system by its inertia ellipsoid (e.g. three Euler angles of the principle axis and three moments of inertia) and $3N - 9$ single particle coordinates. The moments of inertia $I_i$ are related to the coordinates $y_i$ by

$$I_1 = m(y_2^2 + y_3^2), \quad I_2 = m(y_1^2 + y_3^2), \quad I_3 = m(y_1^2 + y_2^2), \quad (4)$$

where $m$ denotes the mass of the particles.

It is interesting to determine those configurations, where the motion of the many-body system may be described in terms of the collective coordinates $y_i$ only. While such motion would be restricted to *some* invariant manifold in phase space it would not obviously be one of those defined by eq. (1). We may however determine invariant manifolds (1) such that the motion on the manifold changes only the inertia ellipsoid of the system and hence may be described entirely by Zickendraht's collective coordinates $y_i$. Two necessary conditions define this situation. First, the number of coordinates on such invariant manifolds may not exceed six. Second, every motion on such invariant manifolds has to change the inertia ellipsoid of the many-body system.

For simplicity let us start with the a system of four particles in two spatial dimensions and the invariant manifold displayed in Fig. 1, i.e.

$$\vec{r}_1 = \begin{bmatrix} x \\ y \end{bmatrix}, \quad \vec{r}_2 = \begin{bmatrix} x \\ -y \end{bmatrix}, \quad \vec{r}_3 = \begin{bmatrix} -x \\ -y \end{bmatrix}, \quad \vec{r}_4 = \begin{bmatrix} -x \\ y \end{bmatrix},$$

and the momenta are chosen by replacing $x \to p_x, y \to p_y$. Computation of the moments of inertia yield the collective Zickendraht coordinates $y_1 = 2x$, $y_2 = 2y$. On the invariant manifold the remaining coordinates are given by $s_{11} = s_{12} = s_{21} = -s_{22} = -s_{31} = -s_{32} = -s_{41} = s_{42} = 1/2$. This shows that every motion on the invariant manifold changes the moments of inertia only and therefore decouples from the single-particle motion.

Consider next the example of an eight-body system in three dimensions. Let

$$\vec{r}_1 = \begin{bmatrix} x \\ y \\ z \end{bmatrix}, \quad \vec{r}_2 = \begin{bmatrix} -y \\ x \\ z \end{bmatrix}, \quad \vec{r}_3 = \begin{bmatrix} -x \\ -y \\ z \end{bmatrix}, \quad \vec{r}_4 = \begin{bmatrix} y \\ -x \\ z \end{bmatrix}, \quad \vec{r}_{4+i} = \vec{r}_i(z \leftrightarrow -z) \quad (5)$$

denote a configuration restricted to the invariant manifold displayed in Fig. 2 (a) with $C_{4h}$ symmetry. (The momenta are chosen by replacing $x \to p_x, y \to p_y, z \to p_z$ in eq.(5).) The moments of inertia are $I_1 = I_2 = 4m(x^2 + y^2) + 8mz^2, I_3 = 8m(x^2 + y^2)$ and yield collective coordinates (4) $y_1^2 = y_2^2 = 4(x^2 + y^2), y_3^2 = 8z^2$. Since the inertia ellipsoid is symmetric we have freedom in the choice of two of its principle axis. Using

$$\vec{y}_1 = 2 \begin{bmatrix} x \\ y \\ 0 \end{bmatrix}, \quad \vec{y}_2 = 2 \begin{bmatrix} -y \\ x \\ 0 \end{bmatrix}, \quad \vec{y}_3 = \sqrt{8} \begin{bmatrix} 0 \\ 0 \\ z \end{bmatrix},$$

one obtains constant single-particles coordinates $s_{11} = -s_{31} = s_{51} = -s_{71} = s_{22} = -s_{42} = s_{62} = -s_{82} = 1/2, s_{13} = s_{23} = s_{33} = s_{43} = -s_{53} = -s_{63} = -s_{73} = -s_{83} = 1/\sqrt{8}$ for the motion on the invariant manifold. Thus, on the invariant manifold the single-particle motion decouples from the collective one. Similar results hold for the six particle configuration displayed in Fig. 2.

It is also instructive to consider a more complex situation. The configuration

$$\vec{r}_1 = \begin{bmatrix} x \\ y \\ z \end{bmatrix}, \quad \vec{r}_2 = \begin{bmatrix} -y \\ x \\ z \end{bmatrix}, \quad \vec{r}_3 = \begin{bmatrix} -x \\ -y \\ z \end{bmatrix}, \quad \vec{r}_4 = \begin{bmatrix} y \\ -x \\ z \end{bmatrix},$$

$$\vec{r}_5 = \begin{bmatrix} x \\ -y \\ -z \end{bmatrix}, \quad \vec{r}_6 = \begin{bmatrix} y \\ x \\ -z \end{bmatrix}, \quad \vec{r}_7 = \begin{bmatrix} -x \\ y \\ -z \end{bmatrix}, \quad \vec{r}_8 = \begin{bmatrix} -y \\ -x \\ -z \end{bmatrix},$$

displays $D_4$ symmetry and differs from configuration (5) by a shearing motion. Like in the previous example, the moments of inertia are given by $I_1 = I_2 = 4m(x^2 + y^2) + 8mz^2, I_3 = 8m(x^2 + y^2)$ and the ellipsoid of inertia is symmetric. However, no choice of the principal axis allows to fulfill eqs. (3) with *constant* single-particle coordinates $s_{ij}$. Therefore, single-particle degrees of freedom depend on collective degrees of freedom and a decoupling does not exist using Zickendraht's coordinate system. However, decoupling is achieved if we use the projector (2) and find coordinates on the invariant manifold and perpendicular to it. The collective motion does not correspond to vibrations or rotations of the inertia ellipsoid only. These findings are interesting e.g. in relation with with the magnetic dipole mode in nuclei [15] since this type of collective behavior is associated with a shearing motion.

## A SIMPLE ILLUSTRATION

Let us consider a system of four particles in two dimensions with quartic one- and two-body interactions. The invariant manifold of Fig. 1 is defined as those points which are invariant under $\mathcal{H} = \{E, \sigma_x P_{(12)(34)}, \sigma_y P_{(14)(23)}, C_2 P_{(13)(24)}\}$. Here $E$ denotes the identity, $P$ a permutation of particles as indicated, $\sigma$ a reflection at the axis indicated, and $C_2$ a rotation about $\pi$. Thus, $\mathcal{H} = C_{2v}$ with four IRs labeled by $\nu = A_1, B_1, A_2, B_2$ [17]. Let $\vec{q} = (x_1, x_2, x_3, x_4, y_1, y_2, y_3, y_4)$ denote a coordinate vector in position space

($x_i, y_i$ denote the coordinates of the $i$'th particle). We have

$$E\vec{q} = (x_1, x_2, x_3, x_4, y_1, y_2, y_3, y_4),$$
$$\sigma_x P_{(12)(34)} \vec{q} = (x_2, x_1, x_4, x_3, -y_2, -y_1, -y_4, -y_3)$$
$$C_2 P_{(13)(24)} \vec{q} = (-x_3, -x_4, -x_1, -x_2, -y_3, -y_4, -y_1, -y_2)$$
$$\sigma_y P_{(14)(23)} \vec{q} = (-x_4, -x_3, -x_2, -x_1, y_4, y_3, y_2, y_1).$$

From the character table of $C_{2v}$ [17] and the projectors (2) one obtains the following basis vectors corresponding to the IR labeled by

$A_1$ : $\quad e'_1 = (1,1,-1,-1,0,0,0,0)/2, \quad e'_2 = (0,0,0,0,1,-1,-1,1)/2,$
$B_1$ : $\quad e'_3 = (1,1,1,1,0,0,0,0)/2, \quad e'_4 = (0,0,0,0,1,-1,1,-1)/2,$
$A_2$ : $\quad e'_5 = (1,-1,-1,1,0,0,0,0)/2, \quad e'_6 = (0,0,0,0,1,1,-1,-1)/2,$
$B_2$ : $\quad e'_7 = (1,-1,1,-1,0,0,0,0)/2, \quad e'_8 = (0,0,0,0,1,1,1,1)/2.$

The vectors associated with the identical IR $A_1$ span the two-dimensional invariant manifold and the vectors associated with the IRs $B_1, A_2, B_2$ span the transverse directions.

Having specified the invariant manifold and appropriate coordinates we now consider the interacting four-body system with the Hamiltonian

$$H = \sum_{i=1}^{4} \left( (p_{x_i}^2 + p_{y_i}^2)/2 + 16(x_i^2 + y_i^2)^2 \right) - \sum_{i<j} \left[ (x_i - x_j)^2 + (y_i - y_j)^2 \right]^2. \quad (6)$$

The stability of the invariant manifold defined above has been studied by computing the full phase space monodromy matrix of several periodic orbits that are inside the invariant manifold [13]. It was found that several orbits are linearly stable in transverse directions or possess rather small transverse stability exponents, while the motion inside the manifold is strongly chaotic.

Let us investigate the decay of collective motion in the corresponding quantum system. To this purpose we consider the time evolution of a Gaussian wave packet $\Psi(\mathbf{r},t)$ that initially is localized on the invariant manifold. The autocorrelation function $C(t) = \langle \Psi(t=0) | \Psi(t) \rangle$ is computed in semiclassical approximation. Within the manifold we used Heller's cellular dynamics [18] which takes into account the nonlinearity of the classical motion. In the transverse direction the time–propagation was done using linearized dynamics only. This approximation neglects any recurrences from the transverse directions and implies a permanent flux of probability out of the manifold and its vicinity. On the time scales considered here, the linearization is justified since the classical return probability to the manifold of transversely escaping trajectories is negligible. It is also important to note that the loss of probability inside the manifold is not severe since the transverse stability exponents are not too large.

We launch wave packets along periodic or aperiodic orbits lying within the invariant manifold and consider their revival as measured by the autocorrelation function. To achieve shorter recurrence times the initial packet was symmetrized with respect to the reflection symmetry of the system within the invariant manifold.

We propagate such wave packets for the invariant manifold. For not too unstable periodic orbits we expect a fairly strong revival after one period, known as the linear revival [18]. As an example, we show in Fig. 3 the real part of the autocorrelation function, calculated for a wave packet on a periodic orbit. We indeed find strong linear revival. However, at larger times we find randomly scattered strong revivals. The revival corresponding to twice the period is not dominant. This implies that a significant fraction of the original amplitude remains near the invariant manifold, and that this fact is not related to the periodic orbit we started on. Revivals calculated for packets started on aperiodic orbits show similar features except for the obvious absence of the linear revival.

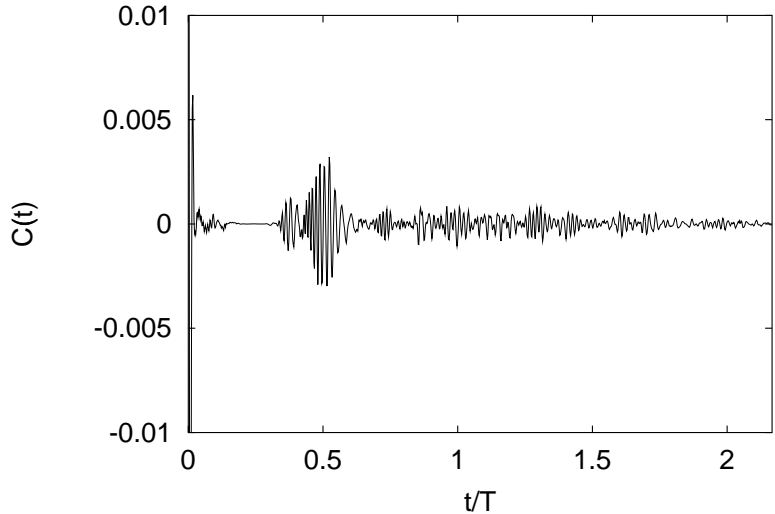

**FIGURE 3.** Autocorrelation function $C(t)$ of a symmetrized wave packet launched on a periodic orbit with period $T$ inside the weakly unstable manifold. In addition to the linear revival around $t = \frac{T}{2}$, a strong nonlinear revival is seen for larger times.

Thus, wave packets may have unusually long life times on certain invariant manifolds characterized by small classical transverse instabilities. This is a quantum and not a classical phenomenon and constitutes an extension of the concept of a scar [19].

## SUMMARY

Low-dimensional invariant manifolds are part of the phase space for interacting many-body systems with permutation and rotation symmetry. We presented coordinates that are naturally adapted to an invariant manifold. There are several configurations of few-body systems, where the motion on the invariant manifold corresponds to a vibration or rotation and may be described in terms of Zickendraht's coordinates, but differs when the collective motion goes beyond that. These results are independent of the details of the Hamiltonian of the $N$-body system, and are entirely determined by rotational and permutational symmetry.

The importance of a given invariant manifold depends on its stability for motion that is close to the manifold. There are few-body systems with invariant manifolds that have vanishing [10, 12] or small [13] instability exponents in transversal directions.

Examples include collinear Helium [10], a chaotic, self-bound three-body system [12] and a four-body systems with anharmonic interactions [13]. In these systems one finds an enhancement in wave function amplitude close to the corresponding invariant manifolds, or large revival probabilities for wave packets that are launched on such manifolds. For low level densities the strong revival can result in individual, strongly scarred states which should be identified with low-lying collective states. For high level densities the strength of these modes will be divided among many individual states thus giving rise to intermediate structure.

We thank T. Guhr for useful discussions and acknowledge financial support by CONACyT project 25192E and DGAPA (UNAM) project IN112200. TP acknowledges support as a Wigner Fellow and staff member at Oak Ridge National Laboratory, managed by UT-Battelle, LLC for the U.S. Department of Energy under contract DE-AC05-00OR22725.

# REFERENCES

1. T. Guhr, H. A. Weidenmüller, Ann. Phys. **193** (1989) 472
2. J. Blocki, F. Brut, T. Srokowski, and W. J. Swiatecki, Nucl. Phys. A **545** (1992) 551c
3. W. D. Heiss, R. G. Nazmitdinov, and S. Radu, Phys. Rev. Lett. **72** (1994) 2351
4. S. Drozdz, S. Nishizaki, and J. Wambach, Phys. Rev. Lett. **72** (1994) 2839
5. W. Bauer, D. McGrew, V. Zelevinski, and P. Schuck, Phys. Rev. Lett. **72** (1994) 3771
6. V. R. Manfredi, L. Salasnich, Int. J. Mod. Phys. E **4** (1995) 625
7. V. Zelevinski, Ann. Rev. Nucl. Part. Sci. **46** (1996) 237
8. T. Papenbrock, Phys. Rev. C **61** (2000) 034602
9. T. Papenbrock and T. H. Seligman, Phys. Lett. A **218** (1996) 229
10. D. Wintgen, K. Richter, and G. Tanner, Chaos **2** (1992) 19
11. T. Prosen, Phys. Lett. A **233** (1997) 332
12. T. Papenbrock and T. Prosen, Phys. Rev. Lett. **84** (2000) 262
13. T. Papenbrock, T. H. Seligman, and H. A. Weidenmüller, Phys. Rev. Lett. **80** (1998) 3057
14. T. Papenbrock and T. H. Seligman, preprint nucl-th/0003049, submitted to J. Phys. A.
15. T. Guhr, H. Diesener, A. Richter, C. W. de Jager, H. de Vries, and P. K. A. de Witts-Huberts, Z. Phys. A **336** (1990) 159
16. W. Zickendraht, J. Math. Phys. **12** (1970) 1663
17. M. Hammermesh, *Group theory and its application to physical problems*, Dover Publications, N.Y., 1989
18. E.J. Heller, J. Chem. Phys. **94** (1991) 2723; M.A. Sepulveda, E.J. Heller, J. Chem. Phys. **101** (1994) 8004
19. E.J. Heller, Phys. Rev. Lett. **53** (1984) 1515